\documentstyle[12pt]{article}
\def\thefootnote{\fnsymbol{footnote}}
\def\be {\begin{eqnarray}}
\def\ee {\end{eqnarray}}
\def\beq {\begin{equation}}
\def\eeq {\end{equation}}

\def\pl {Phys. Lett.}

\def\bi {\begin{itemize}}
\def\ei {\end{itemize}}
\def\ben {\begin{enumerate}}
\def\een {\end{enumerate}}

\begin{document}
\hfill {\bf HYUPT-92/07}

\vskip 0.4in
\centerline{\large\bf Rotation Symmetry and Nonabelian Berry Potential}

\vskip 0.6in
\centerline{Hyun Kyu Lee$^{a}$\footnote{Supported in part by the KOSEF
under Grant No.91-08-00-04 and by Ministry of Education(BSRI-92-231)}
and Mannque Rho$^{b}$}

\vskip 0.4in
\centerline{$^{a}$ {\it Department of Physics, Hanyang University}}
\centerline{\it Seoul 133-791, Korea}

\centerline{$^{b}$ {\it Service de Physique Th\'{e}orique, C.E. Saclay}}
\centerline{\it F-91191 Gif-sur-Yvette, France}

\vskip .6in

\centerline{\bf ABSTRACT}
\vskip .5cm
\noindent
We study the role of
rotational symmetry in the systems where nonabelian Berry potentials emerge as
a result of integrating out fast degrees of freedom. The conserved angular
momentum is constructed in the presence of a non-abelian Berry potential, which
is formulated using Grassmann variables. The
modifications on  conventional angular momentum are discussed in close
analogy with monopole systems. The diatomic molecular system discussed by
Zygelman is found to have a similar structure to that of a non-abelian
$SU(2)$ charge coupled to a 't Hooft - Polyakov monopole. The abelian limit
of the Berry potential in diatomic systems is shown to be the same as $U(1)$
monopole and in a large separation limit, we observe the decoupling
associated with a vanishing field tensor.

\par\vfill

\newpage

\renewcommand{\thefootnote}{\arabic{footnote}}
\subsection*{I. Introduction }
\indent

It is well known\cite{jakman} that the conservation laws of a system coupled to
 a symmetric background gauge field persist in modified forms. The simplest
example is the angular momentum of a system coupled to a Dirac ($U(1)$)
magnetic
monopole. The conserved angular momentum is modified to a sum of mechanical
angular momentum and a contribution from the magnetic field\cite{yang}. A more
interesting phenomenon related to this observation is ``spin-isospin"
transmutation \cite{spin}.

The purpose of this note is to investigate the modification on the conserved
angular momentum due to the presence of a non-abelian Berry potential
\cite{berry,wz}. We are motivated to look into this matter with the
aim to gain an insight that can be applied to systems that are more complex
and where a simple delineation of relevant degrees of freedom is not
readily available. We are in particular thinking of the variety of excitations
that take place in strongly interacting systems such as light- and heavy-quark
baryons which have been suggested to be understandable in terms of hierarchy
of induced gauge structure \cite{lnrz,lnrzbig}. The crucial element is that
the gauge structure is generic depending only on symmetries
and independent of the kinds of interactions involved, be that atomic, nuclear
or elementary particle.

In section II, the theory of non-abelian Berry potentials is formulated
not in the usual matrix form but using Grassmann variables. Grassmanian
variables prove to be easier to work with for non-abelian Berry potentials.
For instance, we can avoid the matrix form for action and gauge invariance
is made more transparent.
In section III, the conditions for the modified angular
momentum are discussed in detail and the case of 't Hooft- Polyakov monopole
will be  discussed as an example. In section IV, the method will be used
to construct conserved angular momenta for a system coupled to Berry
potentials.
Particular attention is given to the case of the diatomic molecule discussed by
Zygelman \cite{zg}. Both the $R\rightarrow 0$ and $R\rightarrow \infty$ limits
for the diatomic molecule are discussed.

\subsection*{II. Non-abelian Berry Potential}
\indent
It is now well known\cite{berry} that in a quantum system, induced gauge fields
naturally appear in the space
of slow variables when the fast variables are integrated out. Hereafter they
will be referred to as Berry potentials.  The Schr\"{o}dinger
equation resulting after fast variables are integrated out
is given by the following form,
\be
-\frac{1}{2m}(\vec{\nabla}_{R}  - i\vec{{\bf A}})^2 {\bf \Psi} =
i \frac{\partial {\bf \Psi}}{\partial t}\label{schrod}
\ee
where $\vec{{\bf A}}$ is defined by
\be
\vec{{\bf A}}_{a,b} = i \langle a,\vec{R}|\vec{\nabla}|b,\vec{R}\rangle.
\label{amn}
\ee
$|a,\vec{R}\rangle$ is a `snap-shot' eigenstate for a given {\it slow
variable} $\vec{R}$, which
is related to the reference state $|a\rangle$ by a unitary  operator
$U(\vec{R})$ such that
\be
|a,\vec{R}\rangle = U(\vec{R})  |a\rangle.\label{eqU}
\ee
Eq.(\ref{schrod}) is a matrix equation where ${\bf \Psi}$ is a column vector
defined in a vector space described by $|a\rangle$.

Following \cite{casal}, we introduce a
Grassmann variable $\theta_a$ for each $|a\rangle$ and
rewrite the Schr\"{o}dinger equation eq.(\ref{schrod}) as
\be
-\frac{1}{2m}((\vec{\nabla}_{R}  - i\vec{{\bf A}}
(\theta,\theta^{\dagger},\vec{R}))^2 {\bf \psi} =
i \frac{\partial {\bf \psi}}{\partial t}.\label{gschrod}
\ee
In the above equation, internal degrees of freedom are considered to be
dynamical degrees of freedom treated classically in the form of anticommuting
coordinates.    Equation (\ref{gschrod}) can be obtained by quantizing
the system described by the following Lagrangian
\be
{\cal L}=\frac{1}{2}m\dot{\vec{R}}^2 + i\theta^{\dagger}_a(\frac{\partial}
{\partial t}-i\vec{A}^{\alpha}T^{\alpha}_{ab}\cdot\dot{\vec{R}})\theta_b
\label{glagran}
\ee
where ${\bf T}^{\alpha}$ is a matrix representation in the vector space of
$|a\rangle$'s for a generator ${\bf {\cal T}}^{\alpha}$ of $U(\vec{R})$,
\be
[{\bf T}^{\alpha},{\bf T}^{\beta}]=i f^{\alpha\beta\gamma}{\bf T}^{\gamma}.
\label{talgeb}
\ee
Following the standard quantization procedure \cite{casal,diracq}, we obtain
the following commutation relations,
\be
[R_i,p_j]=i\delta_{ij}, \ \ \ \ \ \{\theta_a,\theta_b^{\dagger}\}=i\delta_{ab}.
\label{quantiz}
\ee
It is then straightforward to obtain the Hamiltonian
\be
H=\frac{1}{2m}(\vec{p}-\vec{{\bf A}})^2\label{thetah}
\ee
where
\be
\vec{{\bf A}} &=& \vec{A}^{\alpha}I^{\alpha},\nonumber \\
   I^{\alpha} &=& \theta^{\dagger}_a T^{\alpha}_{ab}\theta_b.\label{ialpha}
\ee
Using the commutation relations, it can be verified that
\be
[I^{\alpha},I^{\beta}]=if^{\alpha\beta\gamma}I^{\gamma}.
\ee
The Schr\"{o}dinger equation
\be
H\psi=i\frac{\partial \psi}{\partial t},
\ee
with eq.(\ref{thetah}) leads to eq.(\ref{gschrod}).
It is clear that the Lagrangian, eq.(\ref{glagran}), is
invariant under the gauge transformation
\be
\vec{A}^{\alpha} &\rightarrow& \vec{A}^{\alpha} + f^{\alpha\beta\gamma}
\Lambda^{\beta}\vec{A}^{\gamma} - \vec{\nabla}\Lambda^{\alpha}\label{gaugea},
\\
        \theta_a &\rightarrow& \theta_a -i\Lambda^{\alpha}T^{\alpha}_{ab}
\theta_b.\label{gaugetr}
\ee
It should be noted that  eq.(\ref{gaugetr})
corresponds to the gauge transformation on $|a\rangle$. We should also note
that the Lagrangian (\ref{glagran}) resembles closely the Lagrangian
obtained in \cite{lnrz} for the chiral bag model of baryon structure
when the sea quarks are integrated out.

\subsection*{III. Conserved Angular Momentum}
\indent

Consider a particle coupled to an external gauge field of 't Hooft
 -Polyakov monopole\cite{monotp} with a coupling constant $g$.
The magnetic field is given by
\be
\vec{\bf B} = -\frac{\hat{r}(\hat{r}\cdot{\bf T})}{gr^2}\label{bmag}
\ee
which is obtained from the gauge field $\vec{{\bf A}}$
\footnote{ More precisely it is an asymptotic form of
't Hooft-Polyakov monopole field.},
\be
A^{\alpha}_i &=& \epsilon_{\alpha i j} \frac{r_j}{gr^2},
\label{hooft}\\
\vec{{\bf B}} &=& \vec{\nabla} \times \vec{{\bf A}} -ig[\vec{{\bf A}},
\vec{{\bf A}}].\label{twoform}
\ee
Using the convention described in the previous section, the Hamiltonian of
a particle coupled to a 't Hooft-Polyakov monopole can be written as
\footnote{Hereafter we put $g=1$ for close anology with eq.(\ref{thetah}).}
\be
H &=& \frac{1}{2m}(\vec{p}-\vec{\bf A})^2\nonumber\\
  &=& \frac{1}{2m}\vec{{\bf D}}\cdot \vec{{\bf D}}\label{dhamil}
\ee
where $\vec{{\bf D}}=\vec{p}-\vec{\bf A}$.

It is obvious that the mechanical angular momentum $\vec{L}_m$ of a particle
\be
\vec{L}_m=m\vec{r}\times\dot{\vec{r}}=\vec{r}\times\vec{{\bf D}}
\ee
does not satisfy the $SU(2)$ algebra after canonical quantization in
eq.(\ref{quantiz})
and moreover it cannot be a symmetric operator that commutes with the
Hamiltonian.
The conventional angular momentum, $\vec{L}_o =\vec{r}\times\vec{p}$,
satisfies the usual angular momentum commutation rule.
However it does not commute with the
Hamiltonian and hence cannot be a conserved angular momentum of the
system.  This observation shows us that the construction of a conserved angular
momentum of a system coupled to a topologically nontrivial gauge field is not
a trivial matter.

The conserved angular momentum can be constructed by modifying $\vec{L}_m$ to
\be
\vec{L} = \vec{L}_m + \vec{{\bf Q}},\label{rxd}
\ee
where $ \vec{{\bf Q}}=\vec{Q}^{\alpha}I^{\alpha}$ is to be as follows.
The methods to determine $\vec{{\bf Q}}$ have been discussed in the
literature \cite{jakman,yang}. Here we adopt a rather straightforward
method.
The first condition required for $\vec{{\bf Q}}$ is the consistency condition
that $\vec{L}$ satisfy the $SU(2)$ algebra
\be
[L_i,L_j]=i\epsilon_{ijk}L_k.\label{lll}
\ee
This leads to an equation for $\vec{{\bf Q}}$,
\be
\vec{r}(\vec{r}\cdot\vec{{\bf B}}) + \vec{r}\vec{{\cal D}}\cdot\vec{{\bf Q}}
-\vec{{\cal D}}(\vec{r}\cdot\vec{{\bf Q}})=0.\label{c1}
\ee
where
\be
\vec{{\cal D}}=\vec{\nabla} - i[\vec{{\bf A}},\ \ \ ].\label{covd}
\ee
The second equation is obtained by requiring that $\vec{L}$ commute with H,
\be
[\vec{L},H]=0.\label{c2}
\ee
Eq.(\ref{c2}) can be replaced by a stronger condition
\be
 [L_i, {\bf D}_j]=i\epsilon_{ijk}{\bf D}_k,\label{c3}
\ee
which leads to
\be
{\cal D}_i{\bf Q}_j +\delta_{ij}\vec{r}\cdot\vec{{\bf B}} -
r_i{\bf B}_j=0. \label{c31}
\ee
It is obvious that $\vec{L}$ satisfying eq.(\ref{c3}) or (\ref{c31}) commutes
with the Hamiltonian eq.(\ref{dhamil}).

It is intersting to note that
eq.(\ref{c31}) is the condition for the ``spherically symmetric potential"
discussed by Jackiw\cite{jakiwm}. Here we can verify it in a more
straightforward way
using eq.(\ref{c3}). Moreover the meaning of spherical symmetry becomes clear
from eq.(\ref{c2}).

In the case of  't Hooft-Polyakov monopole, eqs.(\ref{bmag})
and (\ref{hooft}), it can be shown that
\be
\vec{{\bf Q}} = \hat{r}(\hat{r}\cdot{\bf I})\label{phitm}
\ee
satisfies eqs. (\ref{c1}) and (\ref{c31}). After inserting eq.(\ref{phitm})
into eq.(\ref{rxd}), we get
\be
\vec{L} &=& \vec{L}_m + \hat{r}(\hat{r}\cdot{\bf I})\label{ltm0}\\
        &=& \vec{r} \times \vec{p} + \vec{I},\label{ltm}
\ee
where
\be
I_i=\delta_{i\alpha}I^{\alpha}.\label{ii}
\ee
Eq.(\ref{ltm}, \ref{ii}) shows clearly how the isospin-spin transmutation
occurs in a
system where a particle is coupled to a nonabelian monopole.

This analysis can be applied to the abelian $U(1)$ monopole
just by replacing $\hat{r}\cdot\vec{I}$ by -1 in eqs.(\ref{phitm}) and
(\ref{ltm0})\footnote{We are considering a Dirac monopole with $e=g=1$.}. Then
\be
\vec{Q} &=& \hat{r},\label{phidm}\\
\vec{L}    &=& m\vec{r}\times\dot{\vec{r}} - \hat{r}.\label{ldm0}
\ee
One can rewrite eq.(\ref{ldm0}) in a more familiar form seen
in the literature
\be
\vec{L} = \vec{r}\times\vec{p} - \vec{\Sigma},\label{ldm}
\ee
where
\be
\vec{\Sigma} =\left(\frac{(1-\cos\theta)}{\sin\theta}\cos\phi,\,
\frac{(1-\cos\theta)}{\sin\theta}\sin\phi,\, 1\right).
\ee

\subsection*{IV. Rotational Symmetry of Nonabelian Berry Potential}
\indent

In this section we construct conserved angular momentum in a diatomic
molecule in which a Berry potential couples to the dynamics of slow degrees of
freedom, corresponding to the nuclear coordinate $\vec{R}$,
a system that has been  discussed by Zygelman\cite{zg}.

The Berry potential is defined on the space spanned by the electronic
states $\pi(|\Lambda|=1)$ and $\Sigma(\Lambda=0)$, where $\Lambda$'s are
eigenvalues of the third component of the orbital angular momentum of
the electronic
states.  The electronic states responding to slow rotation of $\vec{R}$,
 $U(\vec{R})$ defined by
\be
U(\vec{R})={\rm exp}(-i \phi L_z){\rm exp}(i \theta L_y){\rm exp}(i \phi L_z),
\label{uzg}
\ee
induce a Berry potential of the form
\be
\vec{{\bf A}} &=& i\langle \Lambda_a|U(\vec{R})\vec{\nabla}U(\vec{R})^{\dagger}
|\Lambda_b\rangle\label{zga0}\\
              &=& \frac{{\bf A}_{\theta}}{R} \hat{{\bf \theta}} +
\frac{{\bf A}_{\phi}}{R sin\theta}\hat{{\bf \phi}},
\ee
where
\be
{\bf A}_{\theta}&=&\kappa(R)({\bf T}_y \cos\phi - {\bf T}_x \sin\phi),
\nonumber\\
{\bf A}_{\phi}&=& {\bf T}_z(\cos\theta - 1) - \kappa(R)\sin\theta
({\bf T}_x \cos\phi + {\bf T}_y \sin\phi). \label{aphi}
\ee
$\vec{{\bf T}}'$s are spin-1 representations of the orbital angular momentum
$\vec{L}$ and $\kappa$ measures the transition amplitude between the
$\Sigma$ and $\pi$ states
\be
\kappa(R)=\frac{1}{\sqrt{2}}|\langle\Sigma|L_x-iL_y|\pi\rangle|.\label{kappa}
\ee
The nonvanishing field strength tensor is given by
\be
\vec{{\bf B}}=\frac{F_{\theta\phi}}{R^2 \sin\theta}=-\frac{(1-\kappa^2)}
{R^2}T_z\hat{R}.\label{zgb0}
\ee
Following the procedure described in section II, introducing a Grassmanian
variable for each electronic state and replacing ${\bf T}$ by {\bf I} defined
in eq.(\ref{ialpha}) and quantizing the corresponding Lagrangian, we obtain
the Hamiltonian
\be
H=\frac{1}{2\mu} (\vec{p} - \vec{{\bf A}})^2 \label{hzg0},
\ee
where $\vec{{\bf A}} = \vec{A}^{\alpha} I^{\alpha}$.
It should be noted that the presence of
the constant $\kappa$ which is not quantized that appears in the
Berry potential is a generic feature of nontrivial nonabelian Berry potentials
as can be seen in many examples \cite{gp,lnrz}.

To find a solution of eq.(\ref{c1}) and
eq.(\ref{c31}), it is better to look into the Hamiltonian in detail.
Using the gauge freedom, the Hamiltonian can be rewritten in the  most
symmetric form. This can be done by subtracting a trivial (or pure gauge) part
out of
the Berry potential, which is equivalent to choosing a new gauge such that
\be
\vec{{\bf A}}'&=& V^{\dagger}\vec{{\bf A}}V + iV^{\dagger}\vec{\nabla}V
\label{zgas}\\
        {\bf F}' &=& V^{\dagger}{\bf F}V\label{zgbs}
\ee
where $V$ is an inverse operation of $U$ in eq.(\ref{uzg}), {\it i.e.},
$V = U^{\dagger}$.  Then
\be
{\bf A}_{\theta}' &=& (1-\kappa)(I_x \sin\phi-I_y \cos\phi),\nonumber\\
{\bf A}_{\phi}' &=& (1-\kappa)\{-I_z \sin^2 \theta +
\cos\theta \sin\theta (I_x \cos\phi + I_y \sin\phi )\},\label{zgasf}
\ee
or more compactly
$$\vec{{\bf A}}' = (1-\kappa)\frac{\hat{R} \times \vec{I}}{R^2},$$
and
\be
\vec{{\bf B}}'=-(1-\kappa^2)\frac{\hat{R}(\hat{R}\cdot{\bf I})}
{R^2}.\label{zgbsf}
\ee
It is remarkable that the above Berry potential has the same structure as
the 't Hooft-Polyakov monopole, eq.(\ref{bmag}) and eq.(\ref{hooft}), except
for the different constant factors
$(1-\kappa)$ for vector potential and $(1-\kappa^2)$ for magnetic field.
Because of these two different factors, however, one cannot simply
take eq.(\ref{phitm}) as a solution of (\ref{c31}) for the case of
nonabelian Berry potentials.

Using the following identities derived from eq. (\ref{zgasf}),
\be
\vec{R}\cdot\vec{{\bf A}}' &=& 0,\label{da}\\
\vec{R} \times \vec{{\bf A}}' &=& -(1-\kappa)\{\vec{I} -
(\vec{I}\cdot\hat{R})\hat{R}\},\label{ca}
\ee
the Hamiltonian, eq.(\ref{hzg0}), can be written as
\be
H = -\frac{1}{2\mu R^2}\frac{\partial}{\partial R}R^2\frac{\partial}
{\partial R}
+ \frac{1}{2\mu R^2}(\vec{L}_o + (1-\kappa)\vec{I})^2
- \frac{1}{2\mu R^2}(1-\kappa)^2(\vec{I}\cdot\hat{R})^2\label{hzg1}
\ee
Now one can show that the conserved angular momentum $\vec{L}$ is
\be
\vec{L} &=&  \vec{L}_o + \vec{I},\label{zglf}\\
        &=&  \mu\vec{R}\times\dot{\vec{R}} + \vec{{\bf Q}},\label{zglmf}
\ee
with
\be
\vec{{\bf Q}} = \kappa \vec{I} + (1-\kappa)\hat{R}(\hat{R}\cdot\vec{I}).
\label{zgqf}
\ee
Hence, in terms of the conserved angular momentum $\vec{L}$,
the Hamiltonian becomes
\be
H = -\frac{1}{2\mu R^2}\frac{\partial}{\partial R}R^2\frac{\partial}
{\partial R}
+ \frac{1}{2\mu R^2}(\vec{L} - \kappa\vec{I})^2
- \frac{1}{2\mu R^2}(1-\kappa)^2\label{hzg2}
\ee
where  $(\vec{I}\cdot\hat{R})^2 = 1$ has been used.

It is interesting to see what happens in the two extreme cases of $\kappa = 0$
and $1$. For $\kappa = 1$, the degenerate $\Sigma$
 and $\pi$ states form a representation of the rotation group and hence the
Berry potential (and its field tensor) vanishes or becomes a pure gauge.
Then $\vec{{\bf Q}} = \vec{I}$ and $\vec{L} = \mu\vec{R}\times\dot{\vec{R}} +
\vec{I} $.  Now $\vec{I}$ can be understood as the angular momentum of the
electronic system which is
decoupled from the spectrum. One can also understand this as the restoration
of rotational symmetry in the electronic system. Physically $\kappa \rightarrow
1$ as $R\rightarrow \infty$.

For $\kappa=0$, the
$\Sigma$ and $\pi$ states are completely decoupled and only the
$U(1)$ monopole field can be developed on the $\pi$
states\cite{moody}. Eq.(\ref{zgqf}) becomes identical  to eq.(\ref{phidm}) as
$\kappa$ goes to zero and the Hamiltonian can be written as
\be
H = -\frac{1}{2\mu R^2}\frac{\partial}{\partial R}R^2\frac{\partial}
{\partial R}
+ \frac{1}{2\mu R^2}(\vec{L}\cdot\vec{L} - 1)\label{hzgu1}
\ee
which is a generic form for a system coupled to an $U(1)$ monopole field.
Physically this corresponds to small internuclear distance at which
the $\Sigma$ and $\pi$ states decouple.  Therefore, a truly
nonabelian Berry potential can be obtained
only for $\kappa$ which is not equal to zero or one.

\subsection*{V. Discussion}
\indent

     We  have described a system coupled with nonabelian Berry potentials using
Grassmann variables. The advantage of this formulation is that the systematic
investigation of the symmetry of the systems becomes much easier than in the
formulation with an action in a matrix form. For instance, the canonical
quantization procedure can be easily implemented
and the gauge invariance  is made
transparent.  We have shown that the correct Schr\"{o}dinger equation
can be reproduced after quantization.

     Two constraint equations for a conserved angular momentum have been
obtained in a straightforward manner: One from the condition that the angular
momentum $\vec{L}$ satisfy the $SU(2)$ algebra and the other from the
requirement $[\vec{L}, H]=0$.  It turns out that the second equation in a
stronger form, eq.(\ref{c31}), is equivalent to the condition for a
``spherically  symmetric gauge potential" obtained by Jackiw. Hence our
derivation is a verification of the condition of Jackiw in a different setting
while clarifying the meaning of spherical symmetry.

     We apply those prescriptions to  a specific example, a diatomic molecular
system discussed by Zygelman to reinvestigate the role of rotational symmetry
in Berry potentials.  We note that it has a similar structure to that of a
system of an $SU(2)$ charge coupled to a 't Hooft-Polyakov magnetic monopole
except for two different factors, $(1-\kappa)$  and $(1-\kappa^2)$ for
vector potential and magnetic field respectively, these being a generic feature
of nonabelian Berry potentials.  The conserved angular
momentum is found to have the same form as that of  't Hooft-Polyakov magnetic
monopole.  For $\kappa=0$, the
$\Sigma$ and $\pi$ states are completely decoupled from each other
such that only the
$U(1)$ monopole field can be developed on the $\pi$
states\cite{moody}. We demonstrate this explicitly by showing
that the Hamiltonian becomes
a generic form for a system coupled to an abelian monopole.
For $\kappa=1$, the angular momentum of the electronic system is completely
decoupled from the spectrum. In this limit, the total angular momentum of
the system is the sum of the mechanical angular momentum, $\vec{L}_m$, and
the angular momentum stored in the electronic system, $\vec{I}$, each of which
is separately conserved. This comes about because of the restoration of
rotational symmetry in the
electronic system as the $\Sigma$ and $\pi$ states become degenerate.

As mentioned in the Introduction, we were led to this investigation by
the belief that
the situation in diatomic molecules presents a generic case and a clear
understanding of the Berry potentials figuring there
could lead to an insight into other
more complex situations. For instance, the
decoupling of slow and fast variables, or equivalently the restoration of
rotational symmetry, seems to occur also in such complex
phenomena as excitation spectra of heavy-quark baryons in strong interaction
physics
although it does not seem to take place in light-quark systems \cite{lnrzbig}.
Our hope is that the lesson from the diatomic molecule could provide an
understanding of this difficult problem.

\subsubsection*{Acknowledgments}

We are grateful for continuous discussions with Maciej Nowak and
Ismail Zahed whose critical remarks motivated us to look into the
problem addressed here.

\end{document}